\documentclass[aps,pra,showpacs,twocolumn,eqsecnum]{revtex4}
\usepackage{graphicx}
\usepackage{amsmath,amssymb,amsthm,dsfont,bm}
\usepackage{color}
\usepackage{soul}
\usepackage{ulem}

\usepackage{braket}

\begin{document}
\preprint{}
\title{Entanglement between total intensity and polarization for pairs of coherent states}
\author{Carlos Sanchidri\'an-Vaca} 
\author{Alfredo Luis}
\email{alluis@fis.ucm.es}
\homepage{http://www.ucm.es/info/gioq}
\affiliation{Departamento de \'{O}ptica, Facultad de Ciencias
F\'{\i}sicas, Universidad Complutense, 28040 Madrid, Spain}
\date{\today}

\begin{abstract}
We examine entanglement between number and polarization, or between number and relative phase, for pairs of coherent states and two-mode squeezed vacuum via linear entropy and covariance criteria.  We consider the embedding of the two-mode Hilbert space in a larger space to get a well-defined factorization of the number-phase variables. This can be regarded as a kind of  proto-entanglement that can be extracted and converted into real particle entanglement via feasible experimental procedures. In particular this reveals interesting entanglement properties of pairs of coherent states.
\end{abstract}

\pacs{42.50.Ar,42.50.Dv, 03.65.Ta, 42.50.Xa}

\maketitle
\section{Introduction}

Entanglement is a distinguishing  feature disclosing rather fundamental nonclassical phenomena, as well as a powerful resource for emerging quantum technologies. Entanglement has many facets so it may be convenient to distinguish different types. In this work we distinguish between particle-like versus variable-like entanglement, also referred to as intersystem versus intrasystem, respectively,  as introduced in Ref. \cite{Ai}. This might be related as well with the distinction  between intermode and intramode correlations in quantum metrology \cite{SQ15}. By particle-like we mean entanglement between variables belonging to two physically different system, that is, local observables. By variable-like we mean entanglement between commuting variables of a single  system, without any clear subsystem splitting. 

In this work we focus on variable-like entanglement for pair of coherent states. We show that there is entanglement between total number and relative-phase or polarization variables, revealed by linear entropy and covariance criteria for pure states. Pure states are maximal preparations, so all fluctuations and correlations might be ascribed to a quantum origin. The key point is that entanglement is equivalent to non classicality in the traditional sense of lack of {\it bona fide} phase-space joint distribution \cite{AF82}.  Thus entanglement in pairs of Glauber coherent states is consistent with previous evidences of nonclassical behavior as revealed by negativity of number-phase Wigner functions \cite{JV95}, nonclassical statistics in photon number detection \cite{AL16a}, and unbalanced double homodyne detection \cite{LM17}, as well as other results \cite{LJ04}. This also fits with the idea that entanglement may be a more widespread property than naively expected \cite{TGL10,MK16,FACCA10}.

This analysis faces the difficulty that there is no Hilbert state factorization because of the emergence of a quantum correlation between intensity and polarization not present in the classical realm. So rigorous and complete conclusions can be obtained only after embedding the system space in a larger space to make room so these variables become truly independent having their own Hilbert spaces. 

Therefore, this variable-like entanglement should be probably better described as proto-entanglement, that may be eventually converted into particle entanglement via the coupling of the system space with an ancilla system. This would effectively simulate the above embedding of the system space.
 
We think there is still much to be understood about entanglement, as revealed by classical entanglement \cite{CE} and fruitful emerging connections between classical and quantum optics \cite{EC}. This can shed a lot of light on the quantum to classical borderline  with interesting consequences, both from fundamental perspectives and technological applications. 

 \section{Settings}
 
Our state space is made of two modes of the electromagnetic field described by the  complex amplitude operators $a_{1,2}$ acting on the corresponding component of the product Hilbert space $\mathcal{H} = \mathcal{H}_1 \otimes \mathcal{H}_2$. 

\subsection{Total number and relative phase operators}

The total number operator is $N = a^\dagger_1 a_1 + a^\dagger_2 a_2$. Relative phase variables can be constructed via the Stokes operators \cite{So}
 \begin{equation}
 S_x = a^\dagger_1 a_2   +  a^\dagger_2 a_1,  S_y = i( a^\dagger_2 a_1  - a^\dagger_1 a_2 ),   S_z = a^\dagger_1 a_1   - a^\dagger_2 a_2 ,
 \end{equation}
 that satisfy the commutation relations 
 \begin{equation}
 [N, S_j ]=0, \quad [S_x,S_y]= 2i S_z ,
 \end{equation}
for $j=x,y,z$, and cyclic permutations, respectively. For example a suitable unitary operator exponential of the phase difference can be constructed via a polar decomposition of the $S_\pm$ operators $S_\pm = S_x \pm i S_y$ \cite{LS93}. Alternatively, there is a non unitary solution as product of the corresponding Susskind-Glogower operators $E_1 E^\dagger_2$ with 
\begin{equation}
E_j | n\rangle = |n-1 \rangle_j, \qquad E_j | 0 \rangle_j =0 ,
\end{equation}
where $|n \rangle_j$ represents the photon-number basis in $\mathcal{H}_j$. The Stokes operators are basic for the description of any two-beam interferometer. Naturally, they describe as well quantum  polarization properties \cite{AL16b}. Interference and polarization are isomorphic  phenomena and we may equally refer to one or the other simply depending on whether the modes $a_{1,2}$ represent the same vibration state or orthogonal vibrations.  

In principle we need not specify the concrete form of the relative-phase or polarization operator, and we may refer to it in a simple generic form as $M$, since  we will only use that $[N,M]=0$. Nevertheless, for definiteness we will mainly illustrate the procedure with $M=S_z$.

\subsection{Hilbert-space factorization}

Naturally, entanglement properties are closely related to the Hilbert space structure and the nature of the observables chosen \cite{HSS}.  In our case, the presence of the operator $N$ in the pair of commuting variables suggests the following change of labels in the photon-number basis
\begin{equation}
\label{nm1}
|n, m \rangle = |n_1 =  \frac{n}{2} +m\rangle_1 |n_2 =  \frac{n}{2}-m \rangle_2  , 
\end{equation}
where
\begin{equation}
\label{nm2}
n=n_1+n_2, \qquad m=\frac{n_1-n_2}{2} .
\end{equation}
Note that for each $n$ the range of possible values for $m$ is bounded and ranges as $m=-n/2,-n/2+1,\ldots n/2$. This forces the following split sum of the Hilbert space,
\begin{equation}
\mathcal{H} =\bigoplus_{n=0}^\infty \mathcal{H}_n ,
\end{equation}
where $\mathcal{H}_n$ is the finite-dimensional Hilbert space of dimension $n+1$ spanned by the basis vectors $|n,m\rangle$ with fixed $n$.

\bigskip

This means a rather unnatural link between total energy and relative phase or polarization, which is universal, independent of the field state. This is that the spectrum of any $M$ in general depends on the value of $N$. This is clearly so for the phase-difference operator with eigenvalues  $2\pi/(N+1)$ \cite{LP96}, and this is behind the Heisenberg limit in quantum metrology. There seems to be no fundamental reason for this purely quantum feature without classical analog.  For example, in the classical domain any polarization state is allowed for any field intensity since they are clearly independent degrees of freedom. 

This basic quantum link between total number and relative phase or polarization prevents any Hilbert-state factorization of the form  $\mathcal{H}= \mathcal{H}_N \otimes \mathcal{H}_M$. This has consequences when assessing entanglement of the $N$, $M$ variables since typical entanglement measures are devised for the cases of Hilbert-state factorization, essentially via partial traces, not easily addressed otherwise. Nevertheless, there is an alternative strategy that avoids this difficulty by embedding  $\mathcal{H}_1 \otimes \mathcal{H}_2$  in a larger space making room to remove the link between variables. This will be examined in more detail in Sec. IV.

\subsection{States}

The main state we are going to consider is the product of Glauber coherent states $|\alpha_1 \rangle_1 |\alpha_2 \rangle_2$ as eigenstates of the complex-amplitude operators $a_j |\alpha_j \rangle_j = \alpha_j |\alpha_j \rangle_j $, with, in the number basis
\begin{equation}
| \alpha \rangle = e^{-|\alpha |^2/2} \sum_{n=0}^\infty \frac{\alpha^n}{\sqrt{n!}} | n \rangle.
\end{equation}

Clearly there is no entanglement between the  $a_{1,2}$ variables, nor between any pair of variables derived from them under energy conserving canonical transformations. The scenario is different for total number and relative phase variables as we will see in Sec. III.

To this end it may be illustrative to express $|\alpha_1 \rangle_1 |\alpha_2 \rangle_2$ in the $|n,m \rangle$ basis as 
\begin{equation}
\label{css}
|\alpha_1 \rangle_1 |\alpha_2 \rangle_2 = \sum_{n=0}^\infty \sqrt{p_n} e^{i n \delta} | n, \Omega \rangle ,
\end{equation}
where $| n, \Omega \rangle$ are the SU(2) coherent states \cite{cs}
\begin{eqnarray}
| n, \Omega \rangle & = & \sum_{m=-j}^j \sqrt{\frac{(2j)!}{(j-m)! (j+m)!} }  \nonumber \\ 
& \times  & \sin^{j+m} \frac{\theta}{2} \cos^{j-m} \frac{\theta}{2} e^{i 2 m \phi} |n,m\rangle, 
\end{eqnarray}
and $p_n$ is the probability of having a total photon number $n$, 
\begin{equation}
\label{pncs}
p_n = e^{-\bar{N}} \frac{\bar{N}^n}{n!}, 
\end{equation}
being  $\bar{N} = |\alpha_1 |^2 + |\alpha_2 |^2$ the total mean number of photons. This expression follows after the change of variables 
\begin{equation}
\alpha_1 = r  \sin \frac{\theta}{2} e^{i \delta}e^{i \phi}, \quad \alpha_2 = r \cos  \frac{\theta}{2} e^{i \delta}e^{-i \phi},
\end{equation}
with $r^2 = \bar{N}$.

\bigskip

On the opposite side we may also consider the two-mode squeezed vacuum 
\begin{equation}
\label{tmsv}
| \xi \rangle = \sqrt{1-|\xi |^2} \sum_{n=0}^\infty \xi^n | n\rangle_1 | n \rangle_2 ,
\end{equation}
with  total mean number of photons  $\bar{N} = 2 | \xi |^2 /(1-| \xi |^2)$. In this state there is clearly entanglement in the $a_{1,2}$ modes. On the other hand, in the $|n,m \rangle$ basis we get that $m=0$ for all $n$, suggesting that $M$ does not depend on $N$. Thus we might expect some kind of factorization in the total number and relative phase variables.

We have the following expression in the $|n,m \rangle$ basis  
\begin{equation}
\label{otmsv}
| \xi \rangle = \sqrt{1-|\xi |^2} \sum_{n=0}^\infty \xi^n | 2n,0 \rangle  ,
\end{equation}
and the following probability of having a total photon number $2n$
\begin{equation}
\label{pntsv}
p_{2n} = \frac{2 \bar{N}^n}{\left ( \bar{N}+2 \right)^{n+1}} .
\end{equation}

\section{Entanglement between total number and relative phase}

We examine the entanglement between total number $N$ and half the number difference $M$, this is the $n$, $m$ variables. This is addressed  from the perspective of two different entanglement criteria valid for pure states, linear entropy and covariance. This procedure is completely general and could be equally well applied to any pair of commuting observables $A$ and $B$ in the form $N= A+B$ and $M= A-B$. The interest of choosing number operators lies on their simplicity both for theoretical and practical reasons.

\subsection{Linear entropy}

For pure states entanglement is clearly recognized by taking the trace with respect of one of the subsystems. In the case of factorization such reduced state, say $\rho_R$ is pure, and mixed otherwise. Purity is disclosed by the trace of its square, $\rho_R^2$, so a good measure of entanglement is $S= 1 - \mathrm{tr} (\rho_R^2 )$. For factorized states we have $S=0$  while maximal entanglement corresponds to $S=1$.

Within this scenario  let us address alternatively the partial trace with respect to the total number  variable $N$ or half the number difference $M$. Even if  there is no Hilbert-state splitting $\mathcal{H}_N \otimes \mathcal{H}_M$, we calculate the closest  analog to the idea of subsystem trace allowed in this case in order to find possible entanglement. A more rigorous approach will be provided later via an enlargement of the Hilbert space.

We consider arbitrary pure states, expressed  in the photon-number basis and in the total number and half the number difference as 
\begin{equation}
\label{ps}
| \psi \rangle = \sum_{k,\ell} c_{k,\ell}  |k \rangle_1 | \ell \rangle_2  =\sum_{n=0}^\infty \sum_{m=-\frac{n}{2}}^{\frac{n}{2}} c_{\frac{n}{2}+m, \frac{n}{2}-m}  |n,m\rangle.
\end{equation}


We can restrict the total density matrix $\rho = | \psi \rangle \langle \psi |$  to the subalgebra generated by $N$ making use of  the closest analog to the partial trace with respect total number. This is done by removing all the coherences between different $n,n^\prime$ values, so only is left the minimal information required to computes de statistics of any operator $M$ commuting with $N$. This is 
\begin{equation}
\label{rM}
\rho_M =\sum_{n=0}^\infty \Pi_n \rho \Pi_n = \sum_{n=0}^\infty | \psi_n \rangle \langle \psi_n | ,
\end{equation}
where $\Pi_n$ are orthogonal projectors on the subspaces of definite total photon number $\mathcal{H}_n$ 
\begin{equation}
\Pi_n = \sum_{m=-\frac{n}{2}}^\frac{n}{2} |n,m\rangle \langle n,m | ,
\end{equation}
and $ | \psi_n \rangle = \Pi_n  | \psi \rangle$ are the corresponding unnormalized projections with 
\begin{equation}
\langle \psi_n | \psi_{n^\prime} \rangle = p_n \delta_{n,n^\prime} , \quad p_n = \sum_{m=-\frac{n}{2}}^{\frac{n}{2}} \left  | c_{\frac{n}{2}+m, \frac{n}{2}-m} \right  |^2, 
\end{equation}
and $p_n$ is the probability of having a total photon number $n$. 

\bigskip

Similarly, the closest analog of the partial trace with respect to the phase-like or polarization variables is  
\begin{equation}
\label{rN}
\rho_N =\sum_{m=-\infty}^\infty \Pi_m \rho \Pi_m = \sum_{m=-\infty}^\infty | \psi_m \rangle \langle \psi_m | ,
\end{equation}
where $\Pi_m$ are orthogonal projectors on the subspaces of definite value $m$ of $M=S_z /2$
\begin{equation}
\Pi_m = \sum_{n=2|m|}^\infty |n,m\rangle \langle n,m | ,
\end{equation}
and $ | \psi_m \rangle = \Pi_m  | \psi \rangle$ are the corresponding unnormalized projections with 
\begin{equation}
\langle \psi_m | \psi_{m^\prime} \rangle = p_m \delta_{m,m^\prime} , \quad p_m = \sum_{n= 2|m|}^\infty \left  | c_{\frac{n}{2}+m, \frac{n}{2}-m} \right  |^2 ,
\end{equation}
and $p_n$ is the probability of having a total value of $M$ equals to $m$. 


With this we can compute the corresponding linear entropies for the $\rho_M$ and  $\rho_N$ , leading to
\begin{equation}
\label{le1}
S_M = 1- \sum_{n=0}^\infty p_n^2, \qquad S_N = 1- \sum_{m=-\infty}^\infty p_m^2 .
\end{equation}
Note that in general $S_N \neq S_M$ once again due to the lack of the corresponding Hilbert-space splitting.

\bigskip

In the case of the two-mode coherent state $|\alpha_1 \rangle_1 |\alpha_2 \rangle_2$ we have $p_n$ given by Eq. (\ref{pncs}) 
and then
\begin{equation}
\label{Scs}
S_M = 1- e^{-2 \bar{N}}I_0 \left ( 2 \bar{N} \right ) ,
\end{equation}
where  $I_0$ is the Bessel function of order zero.  The entropy is plotted against  $\bar{N}$  in Fig. 1, showing the entanglement increases as $\bar{N}$ increases with $S_M \rightarrow 1$ when $\bar{N} \rightarrow \infty$.   On the other hand for the range of values examined we have seen numerically that $S_N \simeq S_M$ to the extreme of being indistinguishable.   

\begin{figure}
\begin{center}
\includegraphics[width=6cm]{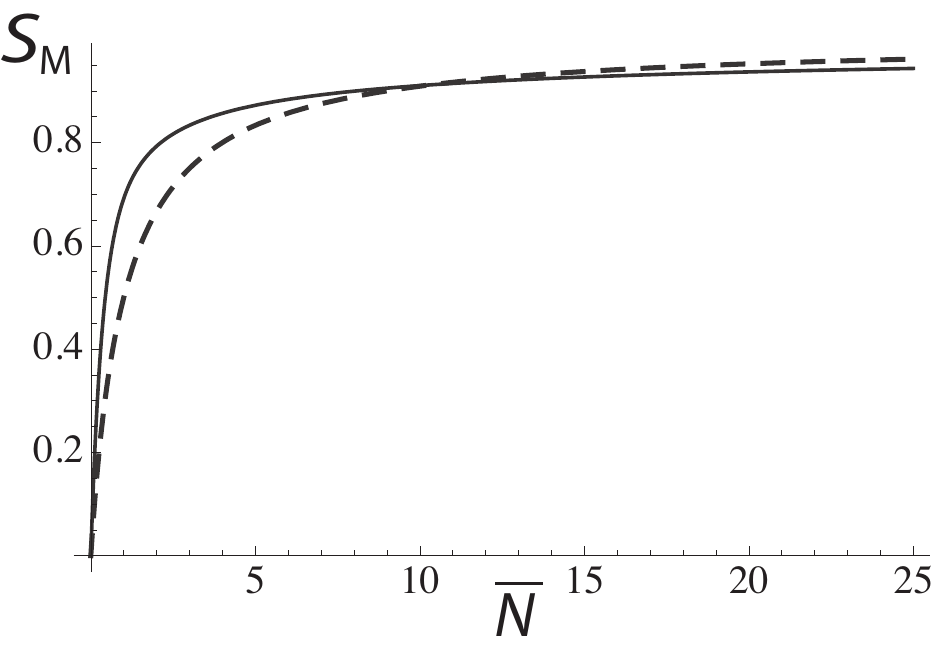}
\end{center}
\caption{Linear entropy $S_M$ versus mean number of photos $\bar{N}$ for a two-mode coherent state (solid line) and a two-mode squeezed vacuum state (dashed line).}
\end{figure}

\bigskip

On the other hand, for the two-mode squeezed vacuum (\ref{tmsv}) we have $p_n$ given by Eq. (\ref{pntsv}) leading to
\begin{equation}
S_M = \frac{\bar{N}}{1+\bar{N}} ,
\end{equation}
with the same behavior than coherent states as shown in Fig. 1, maybe surprisingly. On the other hand, the result is the opposite for $S_N$, since we clearly have that $\rho_N$ is actually a pure state $|\psi_{m=0} \rangle$ so that $S_N = 0$. 

\bigskip

In principle it might be questioned whether these mixedness properly reflect entanglement, after the lack of Hilbert-space factorization. We think that at least they may provide a clear indicator. Maybe a more rigorous and conclusive results will be provided in Sec. IV when embedding the two-mode space in a larger Hilbert space so there is room for the Hilbert-space factorization.

\subsection{Covariance}

Let us now use the covariance criterion studied in Ref. \cite{TGL10}  that establishes that for pure states there is entanglement between $A$ and $B$ variables provided that 
\begin{equation}
\label{AB}
\langle A B \rangle \neq \langle A \rangle \langle B \rangle .
\end{equation}
Clearly this correlation is incompatible with state factorization if $A$ and $B$ act on different spaces. Let us extend this idea to the pair of commuting variables we are considering in this work. Let this time be
\begin{equation}
A=N = a^\dagger_1 a_1 + a^\dagger_2 a_2, \quad B = S_z = a^\dagger_1 a_1 - a^\dagger_2 a_2 .
\end{equation}
So 
\begin{equation}
\langle A B \rangle- \langle A \rangle \langle B \rangle  = \Delta^2 a^\dagger_1 a_1 - \Delta^2 a^\dagger_2 a_2 , 
\end{equation}
so that this criterion predicts entanglement provided that $\Delta^2 a^\dagger_1 a_1 \neq  \Delta^2 a^\dagger_2 a_2$. 

For the pair coherent states $\Delta^2 a^\dagger_j a_j = |\alpha_j |^2$, so that there is entanglement provided that $ | \alpha_1 | \neq  | \alpha_2 |$. Otherwise we  get entanglement by replacing $S_z$ by another Stokes operator. 

\bigskip

On the other hand, for the two-mode squeezed vacuum state (\ref{tmsv}) the equality is always satisfied for all $S_j$ 
\begin{equation}
\langle NS_j \rangle = 0, \qquad \langle S_j \rangle =0 .
\end{equation}
Nevertheless it fails to be satisfied for higher power operators, such as
\begin{equation}
\langle NS_x^2 \rangle = \bar{N} \left (  \bar{N} +2 \right ) \left (  3 \bar{N} +2 \right ) ,
\end{equation}
while
\begin{equation}
\langle S_x^2 \rangle = \bar{N} \left (  \bar{N} +2 \right ) .
\end{equation}

\bigskip

Finally we may consider the rather  odd case $A=B$, where we have entanglement of $A$ with itself provided that $\Delta^2 A \neq 0$. We explain this striking case in Sec. V.

\section{Hilbert-space factorization via embedding}

In principle, the above claims about entanglement may be obscured by the lack of Hilbert-space factorization of the form  $\mathcal{H} = \mathcal{H}_N \otimes \mathcal{H}_M$. This rather technical point can be avoided by extending the state space so that  $\mathcal{H}_1 \otimes \mathcal{H}_2 \subset \mathcal{H}_N \otimes \mathcal{H}_M$   and considering that meaningful states are restricted to some physical sector of $\mathcal{H}_N \otimes \mathcal{H}_M$. Although these observables are non local, they can be measured and tailored theoretically as shown in \cite{nonlocal}

\bigskip

Thus we consider the following embedding of  $\mathcal{H}_1 \otimes \mathcal{H}_2$ into the product of two Hilbert spaces $\mathcal{H}_N \otimes \mathcal{H}_M$ via the following injective correspondence between basis vectors
\begin{equation}
\label{cor}
|n_1 \rangle_1 | n_2 \rangle_2  \rightarrow | n \rangle_N |m \rangle_M   , 
\end{equation}
with
\begin{equation}
n=n_1+n_2, \quad  m=\frac{n_1-n_2}{2}.
\end{equation}
This is essentially the same basis relabeling in Eqs. (\ref{nm1}) and (\ref{nm2}),  but now we admit that there is no restriction on the $n,m$ values in $ | n \rangle_N |m \rangle_M$: $n$ runs over all integers while $m$ runs over all integers and half integers. In other words, the correspondence (\ref{cor}) does not exhausts all basis vectors  $ | n \rangle_N |m \rangle_M$ and there are states $ | n \rangle_N |m \rangle_M$ without preimage. The physical sector holds for basis vector with integer and nonnegative $\frac{n}{2} \pm m$. 

This is essentially a version of previous expansions of the state space to formally include negative numbers in the number basis. This has been used in the quantum phase context to recover the unitarity of the Susskind-Glogower operators, so that $E | 0 \rangle = | - 1 \rangle$ and so on \cite{RGN80}. Then it is said then that physical states are restricted to the sector of nonnegative integers. So it is natural to find that this strategy is also useful in the context dealing also with number-phase variables. 

Now the operation of taking the partial trace with respect to the $N$ variable for a pure state of the form (\ref{ps}) is quite  transparent leading to 
\begin{equation}
\rho_M =\sum_{m,m^\prime=-\infty}^\infty  d_{ m,m^\prime} | m \rangle_M \langle m^\prime | .
\end{equation}
where 
\begin{equation}
\label{dmmp}
d_{m,m^\prime} = \sum_{n=-\infty}^\infty c_{ \frac{n}{2}+m, \frac{n}{2}-m} c^\ast_{\frac{n}{2}+m^\prime,  \frac{n}{2}-m^\prime}. 
\end{equation}
The corresponding trace of the square of $\rho_M$ gives the following linear entropy
\begin{equation}
\label{Se}
S = 1- \sum_{m,m^\prime=-\infty}^\infty  \left | d_{ m,m^\prime} \right |^2.
\end{equation}

Note that in general this entropy is slightly different than the ones in Sec. IIIA  since we are now in a different space. We notice also that now there can be no distinction between the $N$ and $M$ traces. 

\bigskip

Let us compute  $S$ for the pair coherent state  $|\alpha_1 \rangle_1 |\alpha_2 \rangle_2$ with $\alpha_2=0$, this is the vacuum in mode $a_2$. In such a case $c_{k,\ell} = 0$ unless $\ell =0$ that leads to $m=m^\prime = n/2$ in Eq. (\ref{dmmp}) and $d_{m,m^\prime} = d_{n/2,n/2}= p_n$ with the same $p_n$ in Eq. (\ref{pncs}), so that 
\begin{equation}
\rho_M = \sum_{n=0}^\infty e^{-\bar{N}} \frac{\bar{N}^n}{n!} | n/2 \rangle_M \langle n/2 | .
\end{equation} 
Therefore we have the same result in Eq. (\ref{Scs}). However, in this extended case the entropy $S$ no longer depends just on $\bar{N}$ and the result is different under different splittings of the photons between modes. This is for example the case $\alpha_1 = \alpha_2 =1$ for which we have $S = 0.60$. This is below the value obtained for the same state in Sec. III. Besides, the entanglement for ${\mathcal{H}_1 \otimes \mathcal{H}_2}$ is always greater than  in ${\mathcal{H}_N \otimes \mathcal{H}_M}$. This is due to the number of terms in the coefficients in ${\mathcal{H}_1 \otimes \mathcal{H}_2}$ is less than in  ${\mathcal{H}_N \otimes \mathcal{H}_M}$ because of the  embedding. In the physical sector ${\mathcal{H}_N \otimes \mathcal{H}_M}$ there are no terms of the form  $\ket{n}_N\ket{m}_M$ with $|m| > n$. So we have that $S_{\mathcal{H}_1 \otimes \mathcal{H}_2} \geq S_{\mathcal{H}_N \otimes \mathcal{H}_M}$ which means that the restriction to the physical sector reduce the entanglement.

\bigskip

On the other hand, for the two-mode squeezed vacuum (\ref{tmsv}) we get 
\begin{equation}
|\psi \rangle \rightarrow  |\varphi \rangle _N | 0 \rangle_M ,
\end{equation}
where  $ |\varphi \rangle _N$ is an eigenstate of the square of the Susskind-Glogower phase operator in the space $\mathcal{H}_N$,
\begin{equation}
 |\varphi \rangle _N = \sqrt{ 1 - |\xi |^2} \sum_{n=0}^\infty \xi^n | 2n \rangle_M .
 \end{equation}
In this case there is no entanglement and $S=0$ as it could be expected. 

\section{From proto-entanglement to entanglement}

In this Section we present the conversion of proto-entanglement into actual particle entanglement. This is accomplished by the coupling of the system with an appropriate ancilla system. This is a kind of operational  simulation of the system-space embedding discussed in the preceding section.

\subsection{Two-level atom entanglement}

We can provide a very specific and operational scheme considering the nonresonant interaction of the field modes with a pair of two-level atoms. Let  us couple an atom to the variable $N$ and the other one to the variable $M=S_z$  with the following interaction Hamiltonian
\begin{equation}
H_\mathrm{int} = \hbar \lambda \left ( N \sigma_1+ S_z \sigma_2  \right ) ,
\end{equation}
or equivalently 
\begin{equation}
H_\mathrm{int} = \hbar \lambda \left [ a^\dagger_1 a_1  \left ( \sigma_1 +  \sigma_2 \right ) +  
a^\dagger_2 a_2  \left ( \sigma_1 -  \sigma_2  \right ) \right ] ,
\end{equation}
where $\lambda $ is a coupling constant and $\sigma_j =  | e \rangle_j \langle e |$, being $ | g,e \rangle_j$ the corresponding ground and excited states, respectively. This is a really experimentally feasible field-atom coupling as already demonstrated in beautiful experiments via Rydberg interferometry \cite{atc}.

The atoms are initially prepared in the sate $| + \rangle_1 | - \rangle_2$ with 
\begin{equation}
| \pm \rangle_j = \frac{1}{\sqrt{2}} \left ( | g \rangle_j \pm | e \rangle_j \right ) .
\end{equation}
The key point is that when $N$ is even (odd) $S_z$ is also even (odd). We fix the interaction time $\tau$ to be such that $\lambda \tau = \pi$ so that when $N$ and $S_z$ are even the evolution takes the initial state again to the initial state 
\begin{equation}
| + \rangle_1 | - \rangle_2 \longrightarrow | + \rangle_1 | - \rangle_2 ,
\end{equation}
while when $N$ and $S_z$ are odd the evolution produces the following transformation on the atomic state
\begin{equation}
| + \rangle_1 | - \rangle_2 \longrightarrow  | - \rangle_1 | + \rangle_2 .
\end{equation}
Therefore, if the initial state is $| \psi \rangle | + \rangle_1 | - \rangle_2$  the evolved state at time $\tau$ is 
\begin{equation}
\label{tes}
| \psi_e \rangle | + \rangle_1 | - \rangle_2  + | \psi_o \rangle | - \rangle_1 | + \rangle_2 ,
\end{equation}
where $| \psi_{e,o} \rangle$ are the unnormalized orthogonal projections of the initial field state $| \psi \rangle $ on the subspaces of even and odd total photon number $N$, respectively, with 
\begin{equation}
| \psi \rangle  = | \psi_e \rangle + | \psi_o \rangle, \qquad \langle \psi_e  | \psi_o \rangle=0. 
\end{equation}
Note that the state in Eq. (\ref{tes})  is a tripartite entangled state. In order to extract pure particle entanglement in the atomic space we may project the field system on a suitable state. For simplicity this may be the same initial state $| \psi \rangle$, so that the reduced atomic state reads
\begin{equation}
\frac{1}{\sqrt{p_e^2 +p_o^2}} \left ( p_e | + \rangle_1 | - \rangle_2  +  p_o | - \rangle_1 | + \rangle_2 \right ),
\end{equation}
where 
\begin{equation}
p_{e,o}= \langle \psi_{e,o}  | \psi_{e,o} \rangle .
\end{equation}
We can assess the entanglement of this reduced state via a properly normalized version of the linear entropy as $S_a = 2 [1- \mathrm{tr} ( \rho_R^2 )]$ so that it ranges between 0 and 1 with   
\begin{equation}
S_a = 2 \left [ 1- \frac{p_e^4 +p_o^4}{\left ( p_e^2+p_o^2 \right )^2} \right ] .
\end{equation}
We have $S_a =0$ for factorized states $p_e =0$ or $p_o =0$ while $S_a =1$ for maximally entangled states  $p_e = p_o =1/2$ .

In the case of product of coherent states it holds that $p_{e,o}$ depend only on the total mean number of photons $\bar{N}$, as shown by Eqs. (\ref{css}) and (\ref{pncs}) \cite{cs}. More specifically
\begin{equation}
p_e= e^{-\bar{N}} \sinh \bar{N}, \qquad  p_o = e^{-\bar{N}} \cosh \bar{N},
\end{equation}
so that
\begin{equation}
S_a = \tanh^2 \left ( 2 \bar{N} \right ) .
\end{equation}
Moreover, we have that the condition for maximally entangled states is reached actually for very small mean numbers and $p_e \simeq p_o \simeq 1/2$ for $\bar{N}$ as small as $\bar{N} =2$ leading to $S_a = 0.998$. In Fig. 2 we have plotted $S_a$ as a function of $\bar{N}$. Within this same scenario, when the initial field state is the two-mode vacuum  (\ref{tmsv}) we have only even number of photons $p_o=0$, the atomic state always factorizes, and  $S_a =0$. So there is a vey large agreement between this operational entanglement conversion and the above analysis. 

\begin{figure}
\begin{center}
\includegraphics[width=7cm]{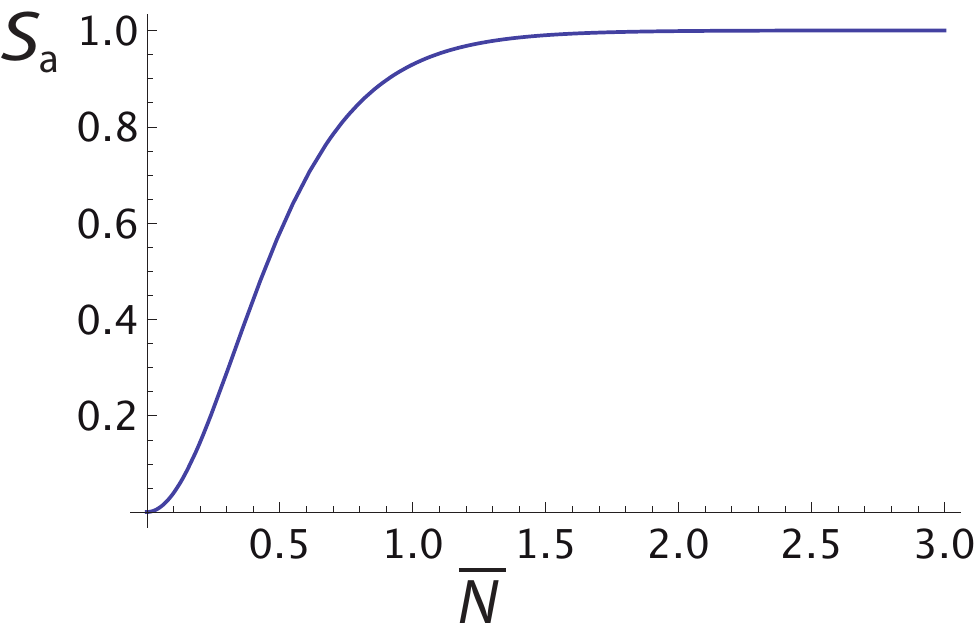}
\end{center}
\caption{The atomic linear entropy $S_a$ versus mean number of photos $\bar{N}$ for a two-mode coherent state.}
\end{figure}

\subsection{Covariance criterion}

Let us show that the proto-entanglement  revealed by  criterion (\ref{AB}) has a complete equivalence with particle entanglement as far one attempts to actually measure both $A$ and $B$ simultaneously to compute $\langle A B \rangle $. To show this we consider that the system space is enlarged from $\mathcal{H}$ to $\mathcal{H} \otimes \mathcal{H}_a$ in order to make room for a joint measurement of $A$ and $B$, always with $[A,B]=0$. For example we consider that $A$ is measured in the original system $\mathcal{H}$ while $B$ is measured in an different system  $\mathcal{H}_a$ prepared in some ancilla state $|\varphi \rangle_a$. In order to transfer information about $B$ from $\mathcal{H}$ to  $\mathcal{H}_a$ we consider the following state transformation 
\begin{equation}
 |\psi \rangle | \varphi \rangle_a \rightarrow |\Psi \rangle =  e^{i B Q_a} |\psi \rangle | \varphi \rangle_a,
 \end{equation}
 where $Q_a$ is some operator acting solely on $\mathcal{H}_a$. Let the observable to be measured in  $\mathcal{H}_a$ be $P_a$ with 
 \begin{equation}
 e^{-i b Q_a}P_a e^{i b Q_a}= P_a + b , \quad   {}_{a} \langle \varphi | P_a | \varphi \rangle_a = 0, 
 \end{equation}
 for every real scalar $b$. 
 
With all this we get 
\begin{equation}
\langle \Psi | AP_a | \Psi \rangle =  \langle \psi | AB | \psi  \rangle ,
\end{equation}
and  
\begin{equation}
\langle \Psi | A| \Psi \rangle =  \langle \psi | A | \psi  \rangle , \quad \langle \Psi | P_a | \Psi \rangle =  \langle \psi | B |  \psi  \rangle .
\end{equation}
Thus, the variable-like entanglement criterion in $| \psi \rangle$ for $A$, $B$ is fully equivalent to the particle-like entanglement criterion in the state $| \Psi \rangle$ for $A$ and $P_a$. 

\bigskip

This particle entanglement may naturally arise even if we measure the same observable twice if $\Delta^2 A \neq 0$. Now we see that there is nothing striking in the entanglement of the state $e^{i A Q_a} |\psi \rangle | \varphi \rangle_a$. 

\bigskip

\section{Conclusions}
We have investigated the entanglement between total number and polarization or relative-phase variables arising in two-mode electromagnetic fields. Since these variables lack a definite splitting of Hilbert spaces this entanglement should be better considered as a kind of  proto-entanglement. Nevertheless we have shown that this  proto-entanglement can be actually extracted and converted into real particle entanglement via feasible experimental procedures. So, this might be regarded as an useful and practical entanglement resource. This is because we have shown that this holds also for products of Glauber coherent states, confirming previous results indicating nonclassical features of coherent states in phase-number variables.

\section*{ACKNOWLEDGMENTS}

A. L. acknowledges financial support from Spanish Ministerio de Econom\'ia y Competitividad 
Project No. FIS2016-75199-P, and from the Comunidad Aut\'onoma de Madrid research  consortium 
QUITEMAD+ Grant No. S2013/ICE-2801.

\end{document}